\documentclass[twocolumn,prb]{revtex4}
\usepackage{graphicx}
\usepackage{dcolumn}
\usepackage{bm}
\usepackage{amsmath}

\begin{document}

\preprint{}
\title{Theory of the tunneling spectroscopy of ferromagnetic superconductors}
\author{T. Yokoyama and Y. Tanaka}
\affiliation{Department of Applied Physics, Nagoya University, Nagoya, 464-8603, Japan%
\\
and CREST, Japan Science and Technology Corporation (JST) Nagoya, 464-8603,
Japan}
%$^2$ Faculty of Science and Technology, University of Twente, 7500 AE,
%Enschede, The Netherlands}
\date{\today}

\begin{abstract}
 We study tunneling conductance in normal metal / insulator / ferromagnetic superconductor junctions. The tunneling spectra show a clear difference  between spin-singlet $s$-wave pairing, spin-triplet opposite spin pairing and spin-triplet equal spin pairing: These pairings exhibit, respectively, gap struture, double peak structure and zero bias peak in the spectra. The obtained result may serve as a tool for determining the pairing symmetry of ferromagnetic superconductors.
\end{abstract}

\maketitle

%--- title ---

%--- author ---

%
%--- address ---

%
%--- date ---

%   Abstract
%-----------------------------------------------------------

%-----------------------------------------------------------

%\section{Introduction}
Magnetism and superconductivity have been under intensive pursuit in the field of low temperature physics. 
Recently the interplay of them has also attracted much attention because nontrivial phenomena are predicted or found experimentally. 
Such  phenomena are expected to occur in ferromanget/superconductor junctions\cite{buzdinrev,Efetov2,golubovrev} and also in ferromangnetic superconductors (FS). Up to now, several bulk materials, e.g., UGe$_2$\cite{Saxena}, ZrZn$_2$\cite{Pfleiderer} and URhGe\cite{Aoki}, are identified as FS. How Cooper pairs are formed in FS or under the coexistence of ferromagnetism and superconductivity is an interesting problem.  However the pairing symmetries of FS are still controversial.

Ferromagnetic superconductors seem to be triplet superconductors because singlet pairing and ferromagnetism are antagonist while triplet pairing have a uniform magnetic moment. However the possibility of $s$-wave pairing cannot be excluded. \cite{Karchev,Singh,Cuoco,Blagoev,Dahal,Kotegawa,Abrikosov,Suhl}
For example it is predicted that UGe$_2$ can have $s$-wave superconductivity mediated by local ferromagnetic spins.\cite{Abrikosov,Suhl}  The study of the nuclear relaxation rate cannot rule out the possibility of $s$-wave pairing in UGe$_2$.\cite{Dahal,Kotegawa} A weak ferromagnetic Fermi liquid theory also suggests the possibility of $s$-wave superconductivity.\cite{Blagoev}
Therefore detailed comparison between theoretical predictions and experimantal data is required to settle this problem. 
Then the properties of thermodynamic quantities  should be noted: For example  equilibrium thermodynamic quantities for Balian-Werthamer state of $p$-wave pairing, which is realized in B phase of $^3$He, are expected to show $s$-wave property because its gap is constant.\cite{Sigrist} In this way, equilibrium thermodynamic quantities for $p$-wave pairing could not be clearly distinguished from those of $s$-wave pairing. Therefore nonequilibrium quantities are more desirable to compare with experimental data. Although some predictions are made on the properties of junctions with equal spin pairing(ESP) FS,\cite{Brataas,Zhao,Gronsleth,Linder} the study of tunneling spectra for possible candidate pairings of FS is insufficient.

Tunneling spectroscopy provides an important information on the superconducting gap and its pairing symmetry. 
In normal metal / supercunductor junctions, Andreev reflection (AR)\cite{Andreev} is a key concept for low energy transport. Blonder, Tinkham and Klapwijk (BTK) formulated  the tunneling conductance where the AR is taken into account \cite{BTK}. This enables us to study the energy gap of superconductors. The generalization of the BTK formula for normal metal / unconventional superconductor junctions are also useful to study the properties of unconventional superconductors \cite{TK,Yamashiro,Yokoyama} because  the tunneling conductance  is sensitive to the pairing symmetry due to the formation of midgap Andreev resonant states\cite{TK,Yamashiro}.

 In the present paper we study the tunneling conductance in normal metal / insulator / ferromangnetic superconductor (N/FS) junctions. The tunneling spectra show a clear difference  between spin-singlet $s$-wave pairing, spin-triplet opposite spin pairing(OSP) and spin-triplet equal spin pairing(ESP). This result may be useful in determining the pairing symmetry of ferromagnetic superconductors.

%-----------------------------------------------------------

%\section{Formulation}

Let us start with an effective Hamiltonian  for the Bogoliubov-de Gennes (BdG) equation. The Hamiltonian   reads
\begin{equation}
\check{H}  = \left( {\begin{array}{*{20}c}
   {\hat H\left( {\bf k} \right)} & {\hat \Delta \left( {\bf k} \right)}  \\
   { - \hat \Delta ^ *  \left( { - {\bf k}} \right)} & { - \hat H^ *  \left( { - {\bf k}} \right)}  \\
\end{array}} \right)
\end{equation}
with
$
\hat H\left( {\bf k} \right) = \xi _k  + {\bf h} \cdot {\bf \sigma }
$
, and $
\hat{ \Delta }\left( {\bf k} \right)  = i\Delta \sigma _y$ for singlet  pairing or $\hat{ \Delta } \left( {\bf k} \right) = \left( {{\bf d}\left( {\bf k} \right) \cdot {\bf \sigma }} \right)i\sigma _y 
$ for triplet pairing. Here $ \xi _k $, ${\bf k}$, ${\bf h}$ and ${\bf \sigma }$ denote  electron band energy measured from the Fermi energy,  electron momentum, applied magnetic field and Pauli matrices respectively. In this paper we consider three types of pairings: spin-singlet $s$-wave pairing, spin-triplet OSP and spin-triplet ESP. 
 OSP and ESP are characterized by the relations ${\bf{h}} \times  {\bf{d}}\left( {\bf{k}} \right) =0$ and
 ${\bf{h}} \cdot {\bf{d}}\left( {\bf{k}} \right)=0$ respectively.\cite{Powell}  

We consider a two dimensional ballistic N/FS junctions at zero temperature.
  The N/FS interface 
located at $x=0$ (along the $y$-axis) has an infinitely
narrow insulating barrier described by the delta function $U(x)=U\delta
(x)$. We first consider OSP. 
The BdG equation reads
\begin{equation}
\check{H}\left( {\begin{array}{*{20}c}
   {\hat { u} _ \pm  }  \\
   {\hat{
 v} _ \pm  }  \\
\end{array}} \right) = E_ \pm  \left( {\begin{array}{*{20}c}
   {\hat{
 u} _ \pm  }  \\
   {\hat{
 v} _ \pm  }  \\
\end{array}} \right)
\end{equation}
 for electron-like quasiparticles, and
\begin{equation}
\check{H}\left( {\begin{array}{*{20}c}
   {\sigma _y \hat{
 v} _ \pm  \sigma _y }  \\
   {\sigma _y \hat{
 u} _ \pm  \sigma _y }  \\
\end{array}} \right) =  - E_ \pm  \left( {\begin{array}{*{20}c}
   {\sigma _y \hat{
 v} _ \pm  \sigma _y }  \\
   {\sigma _y \hat{
 u} _ \pm  \sigma _y }  \\
\end{array}} \right)
\end{equation}
for hole-like quasiparticles, 
with
\begin{equation}
E_ \pm   = \sqrt {\left( {\xi _k } \right)^2  + \left| \Delta  \right|^2 }  \pm \left| {\bf{h}} \right| , \label{eo}
\end{equation}

\begin{equation}
\hat{
 u} _\pm =  u_0^ \pm  \left( {1 \pm {\bf \hat{
 h} } \cdot {\bf \sigma }} \right)/2, 
\end{equation}
\begin{equation}
\hat{
 v} _ \pm=v_0^ \pm  \frac{{\hat{
\Delta } ^\dag  }}{\left| \Delta  \right| }\left( {1 \pm {\bf \hat{
 h} } \cdot {\bf \sigma }} \right)/2,
\end{equation}
\begin{eqnarray}
u_0^ \pm   = \sqrt {\frac{1}{2}\left( {1 + \frac{{\sqrt {E_ \pm  ^2  \mp h - \left| {\Delta ^2 } \right|} }}{{E_ \pm   \mp h}}} \right)}, \label{uo}
\\ \quad v_0^ \pm   = \sqrt {\frac{1}{2}\left( {1 - \frac{{\sqrt {E_ \pm  ^2  \mp h - \left| {\Delta ^2 } \right|} }}{{E_ \pm   \mp h}}} \right)}, \label{vo}
\end{eqnarray}
$
\hat{\bf h} = {\bf h}/\left| {\bf h}\right | 
$ and
 $
\left| \Delta  \right|^2  = \frac{1}{2}Tr\hat{
\Delta } \hat{
\Delta } ^\dag.  
$
We assume $\Delta < h$ because otherwise the gap vanishes for the "-" state as can be seen in Eq. (\ref{eo}). The solution of the BdG equation for $s$-wave pairing have the same form  as that of OSP and obtained by choosing $
\hat{ \Delta }\left( {\bf k} \right) = i\Delta \sigma _y$. 
Below we consider unitary state for triplet superconductors and choose, as a model calculation,  
${\bf{h}} =  - h{\bf{\hat z}}$, $ {\bf{d}}\left( {\bf{k}} \right) = \Delta \left( {k_x  + ik_y } \right)/k{\bf{\hat z}}$ for OSP, and ${\bf{d}}\left( {\bf{k}} \right) = \Delta \left( {k_x  + ik_y } \right)/k{\bf{\hat x}}$ for ESP. Here $\bf{\hat x}$ and $\bf{\hat z}$ are unit vectors oriented to $x$- and $z$-axis respectively. 
For ESP, eigenfunctions for the Hamiltonian are given by 
\begin{equation}
\left( {\begin{array}{*{20}c}
   {u_0^ -  }  \\
   0  \\
   { - v_0^ -  e^{ - i\theta } }  \\
   0  \\
\end{array}} \right),\left( {\begin{array}{*{20}c}
   {v_0^ -  }  \\
   0  \\
   { - u_0^ -  e^{ - i\theta } }  \\
   0  \\
\end{array}} \right),\left( {\begin{array}{*{20}c}
   0  \\
   {u_0^ +  }  \\
   0  \\
   {v_0^ +  e^{ - i\theta } }  \\
\end{array}} \right),\left( {\begin{array}{*{20}c}
   0  \\
   {v_0^ +  }  \\
   0  \\
   {u_0^ +  e^{ - i\theta } }  \\
\end{array}} \right),
\end{equation}

\begin{eqnarray}
u_0^ \pm   = \sqrt {\frac{1}{2}\left( {1 + \frac{{\sqrt {E_ \pm  ^2  - \left| {\Delta ^2 } \right|} }}{{E_ \pm  }}} \right)},\label{ue}
 \\  v_0^ \pm   = \sqrt {\frac{1}{2}\left( {1 - \frac{{\sqrt {E_ \pm  ^2  - \left| {\Delta ^2 } \right|} }}{{E_ \pm  }}} \right)}.\label{ve} \\
E_ \pm   = \sqrt {\left( {\xi _k  \pm \left| {\bf{h}} \right|} \right)^2  + \left| \Delta  \right|^2 } 
\end{eqnarray}
 where $\theta$ is an angle with respect to the interface normal in the N region. 
Note that the magnitude of $h$ can be larger than that of $\Delta$ for ESP because Cooper pairs are insensitive to the exchange field. 
%%%%%%%%%%%%%%%%%%%%%%%%%%%%%%%%%%%%%%%%%%%%%%%%%%%%%%%%%%%%
%%%%%%%%%%%%%%%%%%%%%%%%%%%%%%%%%%%%%%%%%%%%%%%%%%%%%%%%%%%%%

We will calculate the tunnling condunctance, following the BTK method.\cite{BTK,TK} 
Wave function $\psi(x)$  for $x \le 0$ (N region) is represented as 
\begin{widetext}
\begin{equation}
\psi \left( x \le 0 \right) = \left[ {\left( {\begin{array}{*{20}c}
   1  \\
   0  \\
   0  \\
   0  \\
\end{array}} \right)e^{ik_F \cos \theta x}  + a\left( {\begin{array}{*{20}c}
   0  \\
   0  \\
   0  \\
   1  \\
\end{array}} \right)e^{ik_F \cos \theta x} 
   + b\left( {\begin{array}{*{20}c}
   1  \\
   0  \\
   0  \\
   0  \\
\end{array}} \right)e^{ - ik_F \cos \theta x} } \right]e^{ik_F \sin \theta x}
\end{equation}
\end{widetext}
for an injection wave in up spin state, for $s$-wave pairing and OSP. 
$a$ is AR coefficient and  $b$ is normal reflection (NR) coefficient. 
For an injection wave in down spin state and the junction with ESP, wave functions are given in a similar form. 

Similarly for $x \ge 0$ (FS region) $\psi(x)$ is given by the linear conbination of the eigenfunctions. 
  Note that since the translational symmetry holds for the $y$-direction, the momenta parallel to the interface are conserved.

The wave function follows the boundary conditions:
\begin{eqnarray}
 \psi \left( { + 0} \right) = \psi \left( { - 0} \right), \\ 
 \frac{\partial }{{\partial x}}\psi \left( { + 0} \right) - \frac{\partial }{{\partial x}}\psi \left( { - 0} \right) = \frac{{2mU}}{{\hbar ^2 }}\psi \left( { + 0} \right) .
\end{eqnarray}

Applying BTK theory with AR and NR coefficients for electron injections with up and down spin states, we can calculate the angle-resolved dimensionless conductance for OSP represented in the form:
\begin{equation}
\sigma _{S\sigma }  = \frac{{4(4 + Z_\theta ^2 ) + 16\left| {\Gamma _\sigma ^p } \right|^4  - 4Z_\theta ^2 \left| {\Gamma _\sigma ^p \Gamma _\sigma ^m } \right|^2 }}{{\left| {4 + Z_\theta ^2  - Z_\theta ^2 \Gamma _\sigma ^p \Gamma _\sigma ^m } \right|^2 }}, \label{condo}
\end{equation}
\begin{eqnarray}
 \Gamma _\sigma ^p  = \Gamma _\sigma ^{} e^{ - i\theta } ,\Gamma _\sigma ^m  =  - \Gamma _\sigma ^{} e^{ - i\theta },  \\ 
 \Gamma _\sigma ^{}  = \frac{\Delta }{{E + \sigma h + \sqrt {\left( {E + \sigma h} \right)^2  - \left| \Delta  \right|^2 } }},
\end{eqnarray} 
$\sigma=\pm$, $Z_\theta   = \frac{Z}{{\cos \theta }}$, $Z = \frac{{2mU}}{{\hbar ^2 k_F }}$  with quasiparticle energy $E \equiv E_+=E_-$, effective mass $m$, Fermi wavenumber $k_F$ and Fermi energy $E_F$.
 For $s$-wave pairing, the conductance is given by just replacing $\Gamma _\sigma ^p$ and $\Gamma _\sigma ^p$ with $\Gamma _\sigma$ in Eq.(\ref{condo}). 
 We define $\sigma _{N \sigma }$ as the conductance in the normal state which is given by 
\begin{equation}
\sigma _{N\sigma }  = \frac{4}{{4 + Z_\theta ^2 }}.
\end{equation}
The normalized conductance is represented as 
\begin{equation}
\sigma _T  = \frac{{\int_{ - \frac{\pi }{2}}^{\frac{\pi }{2}} {d\theta \cos \theta \left( {\sigma _{S + }  + \sigma _{S - } } \right)} }}{{\int_{ - \frac{\pi }{2}}^{\frac{\pi }{2}} {d\theta \cos \theta \left( {\sigma _{N + }  + \sigma _{N - } } \right)} }}.\label{conde}
\end{equation}

For ESP, the conductances are given by 
\begin{equation*}
\sigma _{S\sigma}  = 4\lambda _\sigma  \times
\end{equation*}
\begin{equation}
\frac{{{\left\{ {Z_\theta ^2  + (\lambda _\sigma   + 1)^2 } \right\} + 4\lambda _\sigma ^{} \left| {\Gamma _{}^p } \right|^2  - \left\{ {Z_\theta ^2  + (\lambda _\sigma   - 1)^2 } \right\}\left| {\Gamma _{}^p \Gamma _{}^m } \right|^2 } }}{{\left| {(\lambda _\sigma   + 1)^2  + Z_\theta ^2  - \left\{ {Z_\theta ^2  + (\lambda _\sigma   - 1)^2 } \right\}\Gamma _{}^p \Gamma _{}^m } \right|^2 }},
\end{equation}
\begin{equation}
 \Gamma _{}^p  = \Gamma _{}^{} e^{ - i\theta } ,\Gamma _{}^m  =  - \Gamma _{}^{} e^{ - i\theta },
\end{equation}
\begin{equation}
 \Gamma  = \frac{\Delta }{{E + \sqrt {E^2  - \left| \Delta  \right|^2 } }},
\end{equation}

\begin{equation}
\sigma _{N\sigma }  = \frac{{4\lambda _\sigma  }}{{\left( {1 + \lambda _\sigma  } \right)^2  + Z_\theta ^2 }}, 
\lambda _ \sigma   = \sqrt {1 - \frac{\sigma h}{{E_F \cos ^2 \theta }}}.
\end{equation}

Note that $\Theta \left( {\theta _C  - \left| \theta  \right|} \right)$ have to be multiplied for $\sigma=+$ (minority spin) in Eq.(\ref{conde}) with $\theta _C  =\cos^{-1}  \sqrt {\frac{U}{{E_F }}}$ because of the mismatch of Fermi surfaces of majority and minority spins.\cite{Yoshida} Here $\Theta(x)$ is the Heaviside step function.

 In the above we choose the same effective mass in N and FS. In most cases the effective mass in N is much smaller than that in FS. However it is expected  that this effect does not change the results qualitatively for large $Z$ ($Z > 1$).\cite{Zutic} Therefore we choose the same effective mass. The inclusion of the difference of  effective masses is straightforward.\cite{Octavio,Zutic}  Although it is known that other characteristics, e.g., the shape of Fermi surfaces should be taken into account in some phase-sensitive tests, \cite{Zutic2} we use a cylindrical Fermi surface in this paper for simplicity because Fermi surface of FS has very complicated structure.\cite{Yates,Settai,Biasini}

%%%%%%%%%%%%%%%%%%%%%%%%%%%%%%%
% CHANGE
%%%%%%%%%%%%%%%%%%%%%%%%%%%%%%%%

%\section{Results}
\begin{figure}[htb]
\begin{center}
\scalebox{0.4}{
\includegraphics[width=21.0cm,clip]{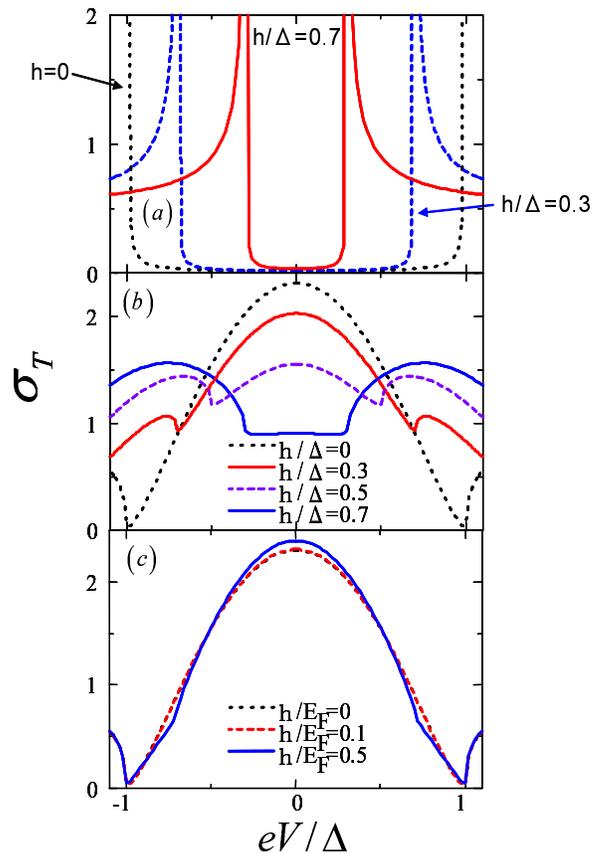}}
\end{center}
\caption{(color online) Normalized tunneling conductance with $Z=10$ for  (a)$s$-wave pairing, (b)OSP and (c)ESP. } \label{f1}
\end{figure}
\begin{figure}[htb]
\begin{center}
\scalebox{0.4}{
\includegraphics[width=19.0cm,clip]{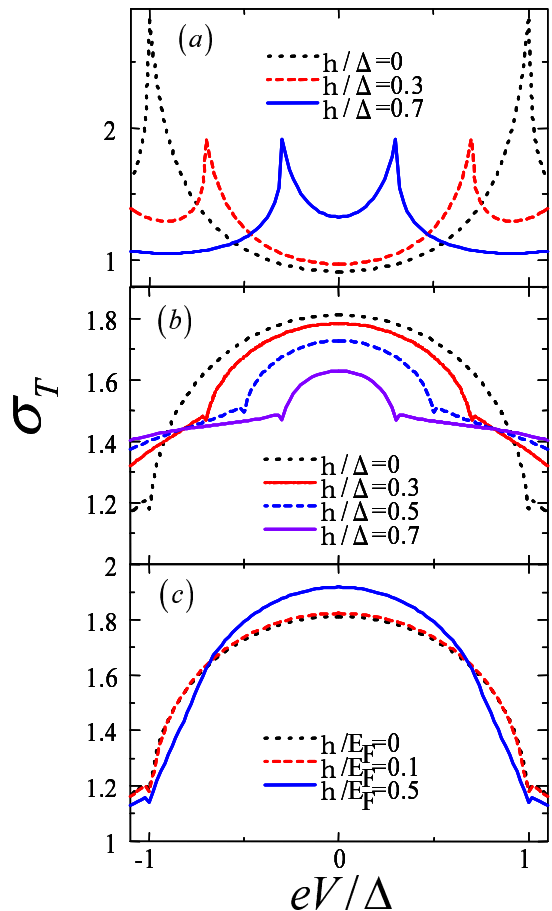}}
\end{center}
\caption{(color online) Normalized tunneling conductance with $Z=1$ for  (a)$s$-wave pairing, (b)OSP and (c)ESP. } \label{f2}
\end{figure}

We study the normalized tunneling conduntace $\sigma _T$ as a function of bias voltage $V$. The conductances with $Z=10$ are shown in Figs. 1(a), 1(b), and 1(c) for $s$-wave pairing, OSP and ESP respectively. 
For $s$-wave pairing, a gap-like sturcture appears at $h=0$.\cite{BTK} With the increase of $h$, the magnitude of the gap is reduced from $2\Delta$ to $2\Delta-2h$ (Fig. \ref{f1} (a)). 
For OSP, a zero bias peak appears at $h=0$ as shown in Fig. 1(b), which stems from the formation of midgap Andreev resonant states.\cite{TK} We find a splitting of peak for OSP as $h$ increases. These shifted structures are attributed to the $h$ dependence of wave function in  Eqs.(\ref{uo}) and (\ref{vo}), and hence expected to emerge for all OSP (not restricted to the present choice of $ {\bf{d}}\left( {\bf{k}} \right)$).   On the other hand, the tunneling conductance has a zero bias peak and is almost independent of the exchange field for ESP as shown in Fig.1 (c). This is because there is no energy shift in the eigenfunctions as shown in Eqs.(\ref{ue}) and (\ref{ve}). We also find that the absence of the shifted structure is expected for all ESP by calculating the eigenfunctions of the Hamiltonian with ESP.  Therefore a clear difference between three types of parings can be seen. Especially when the magnitude of the gap $\Delta$ is comparable to $h$, the tunneling spectra are characterized by gap struture, double peak structure and zero bias peak for $s$-wave pairing, OSP and ESP respectively. 

A corresponding plot for $Z=1$ is shown in Fig. 2. As shown in Fig. 2(a), the reduced dip structure appears for $s$-wave pairing, the width of which is given by  $2\Delta-2h$. When $\Delta \sim h$, the dip transforms into a single peak. As for OSP, a zero bias peak is formed and its width is reduced by the increase of $h$ (see Fig. 2(b)). A zero bias peak remains with the increase of $h$ for ESP as shown in Fig. 2(c).
 Thus there is no qualitative difference between OSP and ESP. This is because the effect of midgap Andreev resonant states becomes weak for small $Z$ and hence the zero bias anomaly is smeared for small $Z$. 
Therefore we find that the difference between for $s$-wave pairing, OSP and ESP becomes clear for large $Z$.

%\section{Conclusions}

In summary we have studied the tunneling conductance in normal metal / insulator / ferromagnetic superconductor junctions. We have found a clear difference in tunneling spectra between spin-singlet $s$-wave pairing, spin-triplet OSP and spin-triplet ESP. The difference is clear for large barrier parameter $Z$. This result may serve as a tool for determining the pairing symmetry of ferromagnetic superconductors.

%=======================================================
%\section{Results}
%=======================================================
%%%%%%%%%%%%%%%%%%%%%%%%%%%%%%%%%%%%%%%%%%%%%%%%%%%
%
%==================================================
T. Y. acknowledges support by the JSPS. 
This work was supported by
NAREGI Nanoscience Project, the Ministry of Education, Culture,
Sports, Science and Technology, Japan, the Core Research for Evolutional
Science and Technology (CREST) of the Japan Science and Technology
Corporation (JST) and a Grant-in-Aid for the 21st Century COE "Frontiers of
Computational Science" . The computational aspect of this work has been
performed at the Research Center for Computational Science, Okazaki National
Research Institutes and the facilities of the Supercomputer Center,
Institute for Solid State Physics, University of Tokyo and the Computer Center.
%======Reference===================================
%

%===============================================================

\end{document}